# INTEGRATED SOLUTION SCHEME FOR HANDOVER LATENCY DIMINUTION IN PROXY MOBILE IPv6

Md. Mahedi Hassan and Poo Kuan Hoong

Faculty of Information Technology, Multimedia University, Cyberjaya, Malaysia
mahedi822002@gmail.com, khpoo@mmu.edu.my

## ABSTRACT

*Recent trends show that there are swift developments and fast convergence of wireless and mobile communication networks with internet services to provide the quality of ubiquitous access to network users. Most of the wireless networks and mobile cellular networks are moving to be all IP based. These networks are connected through the private IP core networks using the TCP/IP protocol or through the Internet. As such, there is room to improve the mobility support through the Internet and support ubiquitous network access by providing seamless handover. This is especially true with the invention of portable mobile and laptop devices that can be connected almost everywhere at any time. However, the recent explosion on the usage of mobile and laptop devices has also generated several issues in terms of performance and quality of service. Nowadays, mobile users demand high quality performance, best quality of services and seamless connections that support real-time application such as audio and video streaming. The goal of this paper is to study the impact and evaluate the mobility management protocols under micro mobility domain on link layer and network layer handover performance. Therefore, this paper proposes an integration solution of network-based mobility management framework, based on Proxy Mobile IPv6, to alleviate handover latency, packet loss and increase throughput and the performance of video transmission when mobile host moves to new network during handover on high speed mobility. Simulations are conducted to analyze the relationship between the network performances with the moving speed of mobile host over mobility protocols. Based on simulation results, we presented and analyzed the results of mobility protocols under intra-domain traffics in micro mobility domain.*

## 1. INTRODUCTION

In recent years, mobile and wireless communications have undergone tremendous changes due to the rapid development in wireless and communication technologies as well as the ever increasing demands by users. Nowadays, mobile end-users are constantly on the go and most of the time, they are moving from one place to another place in rapid pace. As a result, connected mobile devices are also constantly changing their points of attachment to the communication networks, such as Mobile Cellular Networks (MCN), Wireless Local Area Networks (WLAN), Wireless Personal Access Networks (WPAN), and so on. These days, most of the wireless and mobile communication networks are moving towards all IP based. These communication networks are either connected together through the Internet or through private IP core networks. In order to maintain connection, one of the main challenges faced by Mobile Host (MH) is the ability to obtain a new IP address and update its communication partners, while moving amongst these different wireless and mobile networks.

In order to meet the above challenge, Internet Engineering Task Force (IETF) [1] designed a new standard solution for Internet mobility officially called – IPv6 mobility support and popularly named as Mobile IPv6 (MIPv6) [2]. MIPv6 is the modified version of MIPv4, that has great practicality and able to provide seamless connectivity to allow a mobile device to maintain established communication sessions whilst roaming in different parts of the Internet.

When a MH is handed over from one network to another network, it changes the point of attachment from one access router (AR) to another. This is commonly known as *handover* which allows MH to establish a new connection with a new subnet. Handover is also defined as the process of changing between two ARs and when ARs' point of attachment in the network changes. The point of attachment is a base station (BS) for cellular network, or an AR for WLAN. Commonly, handover can be handled in the link layer, if both the ARs are involved in the same network domain. Otherwise, a route change in the IP layer possibly will be needed the so-called network layer handover. In this case, Mobile IPv6 is a standard protocol for handling network layer handover.

For IP-mobility protocols, the IP handover performance is one of the most important issues that need to be addressed. IP handover occurs when a MH changes its network point of attachment from one BS to another. Some of the major problems that may occur during handover are handover latency and packet loss which can degrade the performance and reduce quality of service. In a nutshell, handover latency is the time interval between the last data segment received through the previous access point (AP) and first data segment received through the next AP [3]. The major problem arises with handovers is the blackout period when a MH is not able to receive packets, which causes a high number of packet loss and communication disruption. Such long handover latency might disrupt ongoing communication session and some interruptions. If that change is not performed efficiently, end-to-end transmission delay, jitters and packet loss will occur and this will directly impact and disrupt applications perceived quality of services. For example, handovers that might reach hundreds of milliseconds would not be acceptable for delay-sensitive applications like video streaming and network gaming [3] [4].

Currently, there are several mobility protocols which have been proposed in order to alleviate such performance limitations. One of which is the enhanced version of terminal independent Mobile IP (eTIMIP) [5] [6], which is a kind of mobility management protocol. eTIMIP enhances the terminal independent Mobile IP (TIMIP) [5] by reducing the amount of latency in IP layer mobility management messages exchanged between an MH and its peer entities, and the amount of signaling over the global Internet when a MH traverses within a defined local domain. TIMIP is an example of IP based micro-mobility protocol that allows MH with legacy IP stacks to roam within an IP domain and doesn't require changes to the IP protocol stack of MH in a micro mobility domain.

Compared to the above mobility protocols, Proxy Mobile IPv6 (PMIPv6) [7] defines a domain in which the MH can roam without being aware of any layer 3 ($L_3$) movement since it will always receive the same network prefix in the Router Advertisement (RA). The PMIPv6 specification defines a protocol to support Network-based Localized Mobility Management (NETLMM) [7] [8] where the MH is not involved in the signaling. This new approach is motivated by the cost to modify the protocol stack of all devices to support Mobile IP and potentially its extensions and to support handover mechanisms similar to the ones used in 3GPP/3GPP2 cellular networks.

We make use of Network Simulator, ns-2 [9] in this paper to simulate, examine and compare the performances of eTIMIP, TIMIP, PMIPv6 as well as our proposed integrated solution of PMIPv6 with MIH and Neighbor Discovery (PMIPv6-MIH) in intra-domain traffic with high speed MH. We evaluate the handover latency, packet loss, Peak Signal-to-Noise Ratio (PSNR) [10] and packet delivery throughput of video transmission over PMIPv6 and our proposed integrated solution of PMIPv6-MIH, and also compare the handover latency and packet delivery throughput of transmission control protocol (TCP) and user datagram protocol (UDP) for eTIMIP, TIMIP, PMIPv6 and PMIPv6-MIH in intra-domain traffic.

The rest of this paper is structured as follows: Section 2 briefly explain related research works on the mobility protocols. Section 3 explains overview of media independent handover. Section 4 briefly describes the propose solution scheme. Section 5 shows simulation results of UDP and TCP flow under intra-domain traffic and video transmission over PMIPv6. Finally, Section 6 we conclude the paper and provide possible future works.

## 2. RESEARCH BACKGROUND

For mobility protocols, there are several protocols to reduce handover latency and packet loss, such as the Session Initiation Protocol (SIP) [11] and the Stream Control Transmission Protocol (SCTP) [12]. Both protocols focus on mobility management on an end-to-end basis but they don't have the potential to achieve short handover latency in network layer. The communication sessions in these protocols are initiated and maintained through servers. The behaviour of these protocols is similar to the standard Mobile IP scheme during handovers. However, there are some enhanced Mobile IP schemes that able to reduce the handover latency such as PMIPv6 and CIMS, (Columbia IP Micro-Mobility Suite) [13].

### 2.1. Micro Mobility Protocols

Micro mobility protocols work within an administrative domain which is to ensure that packets are arriving from the internet and addressed to the MHs that forward to the appropriate wireless AP in an efficient manner. It is also called intra-domain traffic [14]. Under the CIMS (Columbia IP Micro-Mobility Suite) project, several micro mobility protocols have been proposed such as –Handoff-Aware Wireless Access Internet Infrastructure (Hawaii) and Cellular IP (CIP).

The CIMS is an extension that offers micro-mobility support. CIMS implements HMIP (Hierarchical Mobile IP) and two micro-mobility protocols for CIP and Hawaii. The CIMS project is mainly focused on intra-domain handover and uses the basic idea of Mobile IP for inter-domain handover.

Subsequently, the CIMS project was enhanced by Pedro et. al. [15] which included the original implementation of TIMIP protocol, and the extended version of TIMIP protocol such as eTIMIP as well as the implementation of CIP, HAWAII, and HMIP protocols. The proposed eTIMIP protocol which is a mobility solution protocol that provides both network and terminal independent mobile architectures based on the usage of overlay micro-mobility architecture.

### 2.2. Enhanced version of Terminal Independent Mobile IP (eTIMIP)

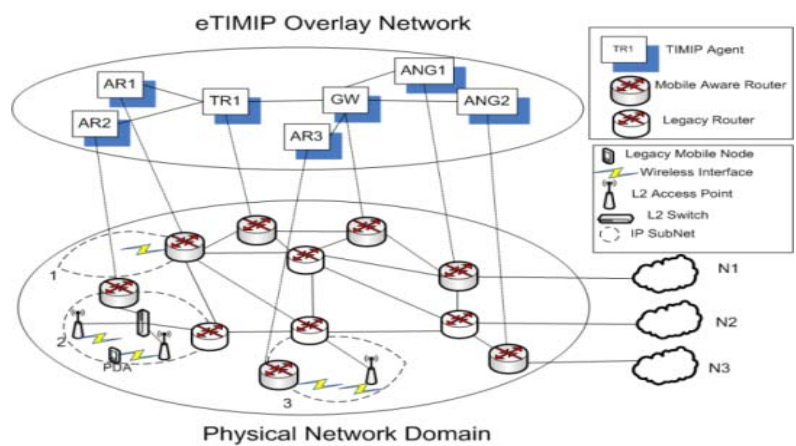

Figure 1. Architecture of eTIMIP

The physical network and overlay network are two complementary networks that are organized in the architecture of eTIMIP. Both networks are separated in the mobile routing from the traditional intra-domain routing which also known as fixed routing. Generally, the physical network can have any possible topology, where it is managed by any specialized fixed routing protocol. The overlay network is used to perform the mobile routing, where it selects routers which support the eTIMIP agents, in which will be organized in a logical tree that supports multiple points of attachment to the external of the domain.

### 2.3. Proxy Mobile IPv6 (PMIPv6)

PMIPv6 is designed to provide an effective network-based mobility management protocol for next generation wireless networks that main provides support to a MH in a topologically localized domain. In general terms, PMIPv6 extends MIPv6 signaling messages and reuse the functionality of HA to support mobility for MH without host involvement. In the network, mobility entities are introduced to track the movement of MH, initiate mobility signaling on behalf of MH and setup the routing state required. The core functional entities in PMIPv6 are the Mobile Access Gateway (MAG) and Local Mobility Anchor (LMA). Typically, MAG runs on the AR. The main role of the MAG is to perform the detection of the MH's movements and initiate mobility-related signaling with the MH's LMA on behalf of the MH. In addition, the MAG establishes a tunnel with the LMA for forwarding the data packets destined to MH and emulates the MH's home network on the access network for each MH. On the other hand, LMA is similar to the HA in MIPv6 but it is the HA of a MH in a PMIPv6 domain. The main role of the LMA is to manage the location of a MH while it moves around within a PMIPv6 domain, and it also includes a binding cache entry for each currently registered MH and also allocates a Home Network Prefix (HNP) to a MH. An overview of PMIPv6 architecture is shown in figure 2.

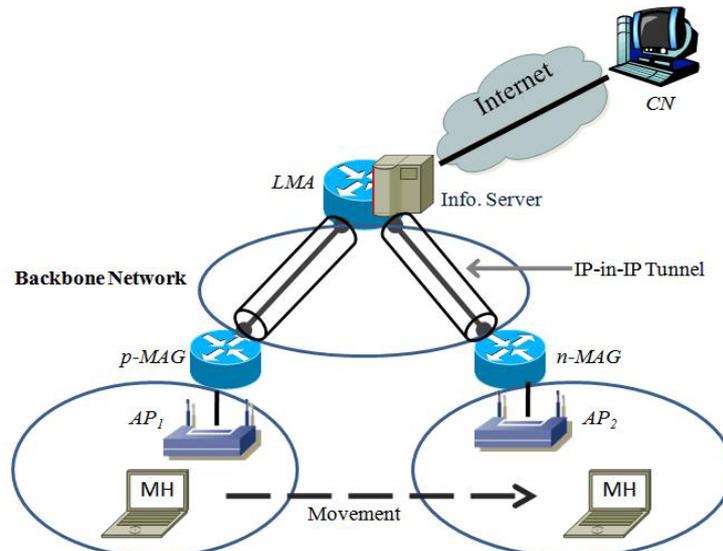

Figure 2. Architecture of PMIPv6

Since the PMIPv6 was only designed to provide local mobility management, it still suffers from a lengthy handover latency and packet loss during the handover process when MH moves to a new network or different technology with a very high speed. Even more, since detecting MHs' detachment and attachment events remains difficult in many wireless networks, increase handover latency and in-fly packets will certainly be dropped at new MAG (n-MAG).

## 3. OVERVIEW OF MEDIA INDEPENDENT HANDOVER

The working group of IEEE 802.21 [15] developed a standard specification, called Media Independent Handover (MIH), which defines extensible media access independent mechanisms that facilitate handover optimization between heterogeneous IEEE 802 systems such as handover of IP sessions from one layer 2 ($L_2$) access technologies to another. The MIH services introduce various signaling, particularly for handover initiation and preparation and to help enhance the handover performance.

The MIH services introduce various signaling, particularly for handover initiation and preparation and to help enhance the handover performance. Figure 3 shows the overall framework of MIH.

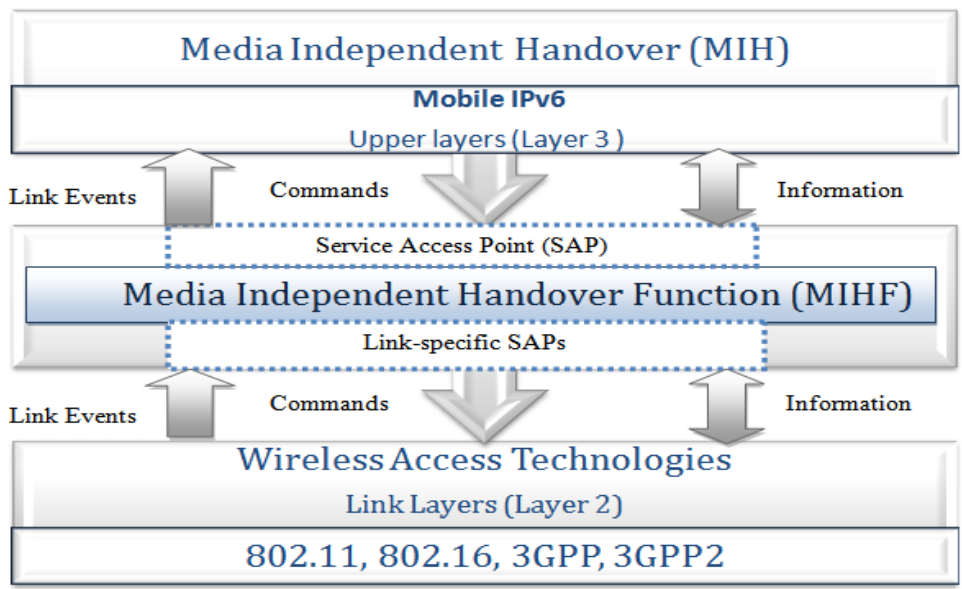

Figure 3. The framework of MIH services

Basically, IEEE 802.21 introduces three different types of communications with different associated semantics [16], the so-called MIH services: Media Independent Event Service (MIES), Media Independent Command Service (MICS) and Media Independent Information Service (MIIS).

MIES introduces event services that provide event classification, event filtering and event reporting corresponding to dynamic changes in link characteristics, links status, and link quality. It also helps to notify the MIH users (MIHU) such as PMIPv6 about events happening at the lower layers like link down, link up, link going down, link parameters report and link detected etc and essentially work as $L_2$ triggers.

MICS provides the command services that enable the MIH users to manage and control link behavior relevant to handovers and mobility, such as force change or handover of an interface. The commands generally carry the upper layers like $L_3$ decisions to the lower layers like $L_2$ on the local device entity or at the remote entity. There are several examples of MICS commands, such as MIH scan, MIH configure, MIH handover initiate, MIH Handover prepare and MIH handover complete.

MIH provides information about the characteristics and services through a MIIS which enables effective handover decisions and system access based on the information about all networks from any single $L_2$ networks. MIIS provides registered MIH users with the knowledge base of the network and information elements and corresponding query-response mechanisms for the transfer of information. By utilizing these services, the MIH users are able to enhance handover performance such as through informed early decisions and signaling. MIIS are classified into three groups, namely general or access network specific information, Point of Attachment (PoA) specific information and vendor specific information.

## 4. PROPOSED INTEGRATED SOLUTION

In response to the PMIPv6 problems mentioned in Section 2, we proposed solution scheme that provides an integrated solution with integrate the analysis of handover latency introduced by PMIPv6 with the seamless handover solution used by MIH as well as the Neighbor Discovery message of IPv6 to reduce handover latency and packet loss on network layer at n-MAG to avoid the on-the-fly packet loss during the handover process. Figure 4 represents the proposed integrated solution of PMIPv6-MIH.

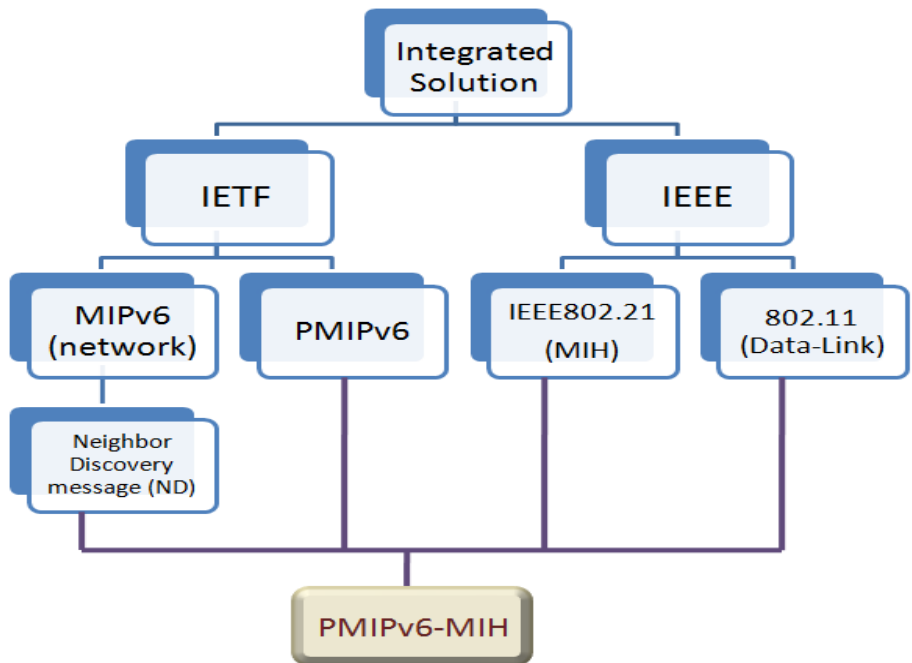

Figure 4. Proposed Integrated Solution

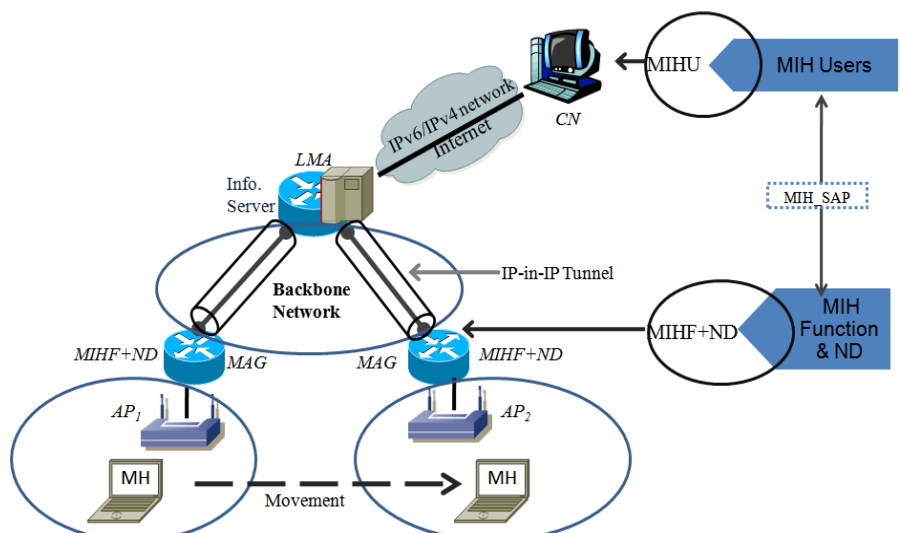

Figure 5. Integrated solution architecture of PMIPv6

Figure 5 presents the key functionality is provided by Media Independent Handover (MIH) which is communication among the various wireless layers and the IP layer. The working group of IEEE 802.21 introduces a Media Independent Handover Function (MIHF) that is situated in the protocol stack between the wireless access technologies at lower layer and IP at upper layer. It also provides the services to the $L_3$ and $L_2$ through well defined Service Access Points (SAPs) [16].

### 4.1. Neighbor Discovery

Neighbor Discovery (ND) enables the network discovery and selection process by sending network information to the neighbor MAG before handover that can helps to eliminate the need for MAG to acquire the MH-profile from the policy server/AAA whenever a MH performs handover between two networks in micro mobility domain. It avoids the packet loss of on-the-fly packet which is routed between the LMA and previous MAG (p-MAG). This network information could include information about router discovery, parameter discovery, MH-profile which contains the MH-Identifier, MH home network prefix, LMA address (LMAA), MIH handover messages etc., of nearby network links.

### 4.2. Analysis of Handover Latency and Assumptions

The overall handover latency consists of the $L_2$ and $L_3$ operations. The handover latency is consequent on the processing time involved in each step of handover procedure on each layer.

The handover latency ($L_{seamless}$) can be expressed as:
$$Lseamless = L_{L_2} + L_{L_3} \quad \ldots\ldots\ldots (1)$$

where $L_{L3}$ represents the network layer as example switching latency and $L_{L2}$ represents link layer as example switching time.

On $L_3$, the handover latency is affected by IP connectivity latency. The IP connectivity latency results from the time for movement detection (MD), configure a new CoA (care-of-address), Duplicate Address Detection (DAD) and binding registration. Therefore, $L_3$ can be denoted as follows:
$$L_3 = T_{config} + T_{DAD} + T_{reg} + T_{move} \quad \ldots\ldots\ldots (2)$$

where $T_{move}$ represents the time required for the MH to receive beacons from n-MAG, after disconnecting from the p-MAG. In order to estimate the movement detection delay, based on the assumptions of mobility management protocols that the times taken for MD are RS and RA messages as follow:
$$T_{move} = T_{RS} + T_{RA} \quad \ldots\ldots\ldots (3)$$

$T_{conf}$ represents the time that taken for new CoA configuration. $T_{reg}$ represents the time elapsed between the sending of Proxy Binding Update (PBU) from the MAG to the LMA and Proxy Binding Advertisement (PBA) from the LMA to the MAG and the arrival/transmission of the first packet through the n-MAG. Binding registration is the sum of the round trip time between MAG and LMA and the processing time as follows:
$$T_{reg} = T_{PBU} + T_{PBA} \quad \ldots\ldots\ldots (4)$$

$T_{DAD}$ represents the time required to recognize the uniqueness of an IPv6 address. Once the MH discovers a new router and creates a new CoA it tries to find out if the particular address is unique. This process is called DAD and it is a significant part of the whole IPv6 process.

As simplification of (2), (3) and (4) equations, it can be expressed as:
$$L_3 = T_{config} + T_{DAD} + T_{PBU} + T_{PBA} + T_{RS} + T_{RA} \quad \ldots\ldots(5)$$

On $L_2$, MH has to perform three operations during the IEEE 802.11 handover procedure such as scanning ($T_{scan}$), authentication ($T_{AAA}$) and re-association ($T_{re-ass}$). Handover latency at $L_2$ can be denoted as follows:

$$L_2 = T_{scan} + T_{AAA} + T_{re-ass} \quad \text{…………………...…..(6)}$$

$T_{scan}$ represents the time that taken the MH performs a channel scanning to find the potential APs to associate with. When MH detects link deterioration, it starts scanning on each channel finding the best channel based on the Received Signal Strength Indicator (RSSI) value.

$T_{AAA}$ represents the time taken for authentication procedure that depends on the type of authentication in use. The authentication time is round trip time between MH and AP.

While $T_{re-ass}$ represents the time needed for re-association consists of re-association request and reply message exchange between MH and AP if authentication operation is successful.

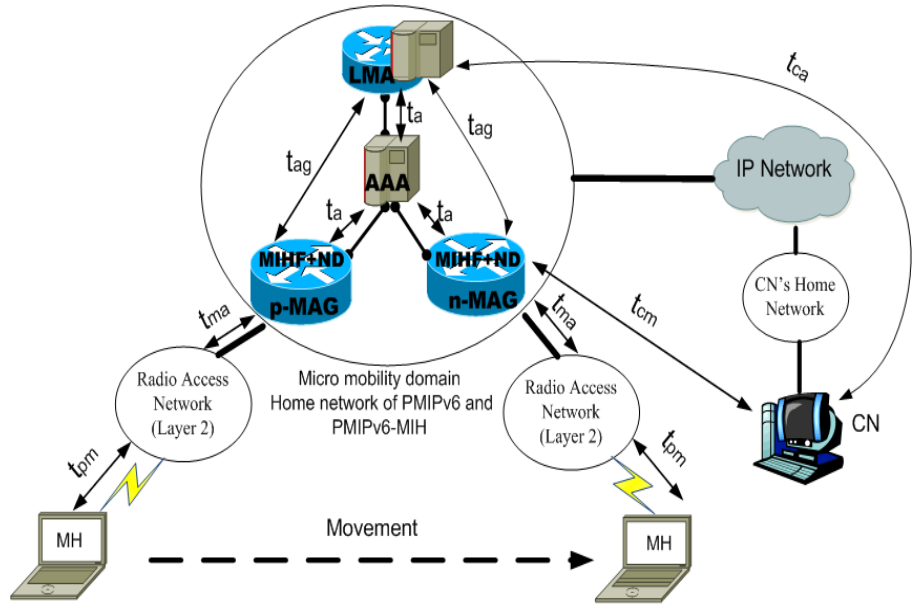

Figure 6. An Analytical Model of PMIPv6 & Integrated solution of PMIPv6-MIH

The following notations are depicted in figure 6 for PMIPv6 & integrated solution of PMIPv6-MIH.

- The delay between the MH and AP is $t_{pm}$, which is required the time for a packet send between the MH and AP through a wireless link.
- The delay between the AP and n-MAG is $t_{ma}$, which is the time between the AP and the n-MAG connected to the AP.
- The delay between the n-MAG and LMA is $t_{ag}$.
- The delay between the LMA and Corresponding Node (CN) is $t_{ca}$.
- The delay between the n-MAG and CN is $t_{cm}$, which is the time required for a packet to be sent between the n-MAG and the CN.
- The delay between the mobility agents and AAA is $t_a$.

As shown in figure 6, we propose an integrated solution of PMIPv6 with MIH and ND to reduce handover latency as the time taken for scanning by informing the MH about the channel information of next APs and use ND message of IPv6 to reduce handover delay and packet loss on network layer at n-MAG to avoid the on-the-fly packet loss during the handover process. Therefore, the handover delay of PMIPv6 and PMIPv6-MIH in a micro mobility domain as follows:

*Registration delay:*
$$T_{PBU} = t_{ag}$$
$$T_{PBA} = t_{ag}$$

We can add the above equations,
$$T_{PBU} + T_{PBA} = 2t_{ag}$$

*Movement Detection delay:*
$$T_{RA} = t_{pm} + t_{ma}$$

*Authentication delay:*
$$T_{AAA} = 2(\text{query} + \text{reply}) = 2(t_a + t_a) = 4t_a$$

*Attachment notification delay:*
The attachment notification delay due to the packet from the AP that informs the MAG of an MH's attachment denote $T_{attach}$

*Configuration and DAD delay:*
PMIPv6 does not require $T_{config}$ and $T_{DAD}$ because MH is already in the PMIPv6 domain. Once the MH has entered and is roaming inside the PMIPv6 domain, $T_{config}$ is not relevant since according to the PMIPv6 specification, the MH continues to use the same address configuration. A $T_{DAD}$ is required for a link-local address since address collision is possible between MH, MAG and all MHs attached to the same MAG. The $T_{DAD}$ may significantly increase handover delay and is a very time consuming procedure. Typically, $T_{DAD}$ is around one second and sometimes can be much more than that. Therefore, PMIPv6 introduces a per-MH prefix model in which every MH is assigned a unique HNP. This approach may guarantee address uniqueness. The new IP address configuration and the DAD operation for global address are appreciable only when the MH first enters the PMIPv6 domain or move to new PMIPv6 domain.

On $L_3$ and $L_2$ equations, Handover delay in PMIPv6 in a micro mobility domain can be expressed as:

$$L_{3_{PMIPv\ 6}} = T_{attach} + 2t_{ag} + t_{pm} + t_{ma} \quad \ldots\ldots\ldots (7)$$

$$L_{2_{PMIPv\ 6}} = T_{scan} + 4t_a + T_{re-ass} \quad \ldots\ldots\ldots (8)$$

During the IEEE802.11 handover procedure the MH performs scanning on the certain number of channels to find the potential APs to associate with. By informing the MH about the channel information of next APs can significantly reduce the scanning time.

However, the scanning time also depends on the type of scanning is used. There are two types of scanning which are defined as active and passive. In active scan mode, MH sends probe request and receives probe response if any AP is available on certain channel. While in passive scan mode, each MHs listens the channel for possible beacon messages which are periodically generated by APs. The handover delay in active scan mode is usually less than in passive scan mode. The operation of passive scan mode depends on the period of beacon generation interval. Therefore, this can provide better battery saving than active scan mode of operation.

As in $L_2$ trigger, the p-MAG has already authenticated the MH and sends the MH's profile which contains MH-Identifier to the n-MAG through the ND message since the MH is already in the PMIPv6 domain and receiving as well as sending information to CN before the handover.

Hence, the authentication delay is eliminated during actual handover. Thus, the $L_2$ handover delay can be expressed as:

$$L_{2\,PMIPv\,6-MIH} = T_{scan} + T_{re-ass} \qquad \ldots\ldots\ldots\ldots\ldots (9)$$

As the parts of $L_3$ handover delay that should be taken into consideration in PMIPv6. Since we proposed the integrated solution of PMIPv6 with MIH services and ND, the number of handover operations should not be considered for overall handover latency. As a result, $L_3$ handover delay is considered only two things in integrated solution of PMIPv6-MIH in a micro mobility domain.

- o When MH attaches to the n-MAG and delivers event notification of MIH_Link_up indication, n-MAG sends a PBU message to the LMA for updating the lifetime entry in the binding cache table of the LMA and triggering transmission of buffer data for the MH
- o RA message

Therefore, the overall handover delay at $L_3$ can be expressed as:

$$L_{3\,PMIPv\,6-MIH} = T_{PBU} + T_{RA} \qquad \ldots\ldots\ldots (10)$$

Based on Analytical model the equations (9) and (10) can be expressed as:

$$L_{2\,PMIPv\,6-MIH} = T_{scan} + T_{re-ass}$$

$$L_{3\,PMIPv\,6-MIH} = T_{PBU} + T_{RA} = t_{ag} + t_{pm} + t_{ma}$$

Seamless Handover Latency of PMIPv6 and integrated solution of PMIPv6 with MIH+ND can be expressed as:

$$L_{seamless\ (PMIPv\ 6)} = L_{L_3\,PMIPv\,6} + L_{L_2\,PMIPv\,6}$$

$$L_{seamless\ (PMIPv\ 6)} = T_{attach} + 2t_{ag} + t_{pm} + t_{ma} + T_{scan} + 4t_a + T_{re-ass}$$

$$L_{seamless\ (PMIPv\ 6-MIH\ )} = L_{L_3\,PMIPv\,6-MIH} + L_{L_2\,PMIPv\,6-MIH}$$

$$L_{seamless\ (PMIPv\ 6-MIH\ )} = t_{ag} + t_{pm} + t_{ma} + T_{scan} + T_{re-ass}$$

### 4.3. Integrated Solution Scheme

The handover latency is mainly caused by the authentication procedure, attachment notification, obtain MH profile and reconfiguration of the default router when the MH access to a new access network. Therefore, we use MIH and Neighbor Discovery message of IPv6 in order to reduce the handover latency, packet loss and increase throughput.

Figure 7 depicts the control and signaling data flow of the proposed integrated solution scheme of PMIPv6-MIH.

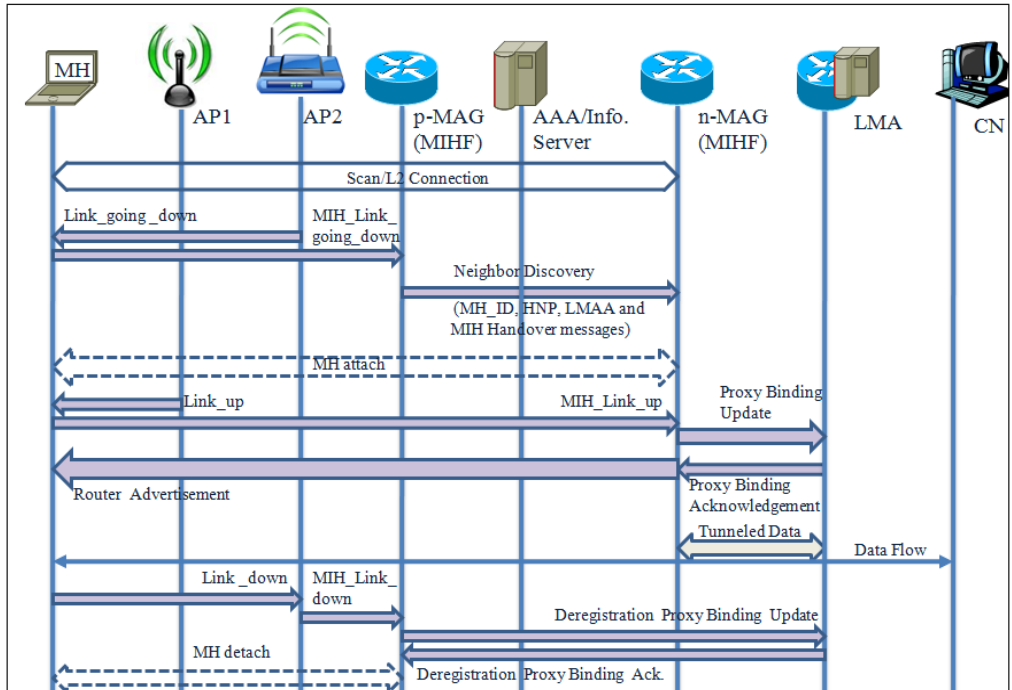

Figure 7. Control and Signaling Data Flow of PMIPv6-MIH

As shown in figure 7, when p-MAG receives a MIH_LGD (Link_going_down) trigger, it transmits the MH profile to its n-MAG using the Neighbor Discovery message of IPv6. This Neighbor Discovery message contains the MH profile, including the MH HNP, MH-ID, and LMAA (LMA address) and MIH handover messages. This eliminates the need for the MAG to acquire the MH profile from the policy server/AAA server whenever an MH performs a handover. The p-MAG starts buffering the packets for the MH in order to avoid the on-the-fly packet loss.

During MH's attachment to the n-MAG, A MIH_Link_up event will be delivered triggering PBU transmission to LMA. After reception of PBU, LMA updates the lifetime for the MH's entry in binding cache table and starts transmission of the buffered data through the tunnel between LMA and MAG.

## 5. SIMULATION EXPERIMENTS AND RESULTS

In order to examine, evaluate and compare the impact on micro mobility domain handover performance of the proposed integrated solution, simulations were performed to compare and evaluate micro mobility protocols by using the ns-2 [9]. The simulation results are conducted in two ways:

1. For the first simulation, two important performance indicators are measured which are the throughput for packet delivery and handover latency. In order to obtain reasonable results, we measure the performance for micro mobility protocol in intra-domain traffic for both TCP and UDP packet flow.

2. While the second simulation results are conducted in video transmission under four important performance metrics that are the Handover Latency, Throughput, Packet Loss and the quality of video streamed which is measured by PSNR. For the PSNR measurements, tools provided by Chih-Heng Ke [17] are used to simulate video transmission simulation over wireless network in ns-2. For our simulations, we utilized the tools of video transmission and used a video stream of MPEG-coded with MPEG4 type frames that is used as a source model for MPEG4 traffic. There are two format sizes such as QCIF (l76 x 144) and CIF (352 x288), where video frame size is the only difference. For our simulations, we convert a video clip file named "Transformers 2" in CIF format and measure the performance for network-based mobility management protocols in a micro mobility domain with high speed movement of the MH.

### 5.1. Simulation Scenario Setup

The simulation scenario setup was implemented as a network-based mobility management solution in the simulation of mobility across overlapping wireless access networks in micro mobility domain. The proposed integrated solution scenario setup is the same as the PMIPv6 but further incorporates MIH functionality in the MH and the MAGs. Thus, the simulation setup scenario is as shown in figure 8 below:

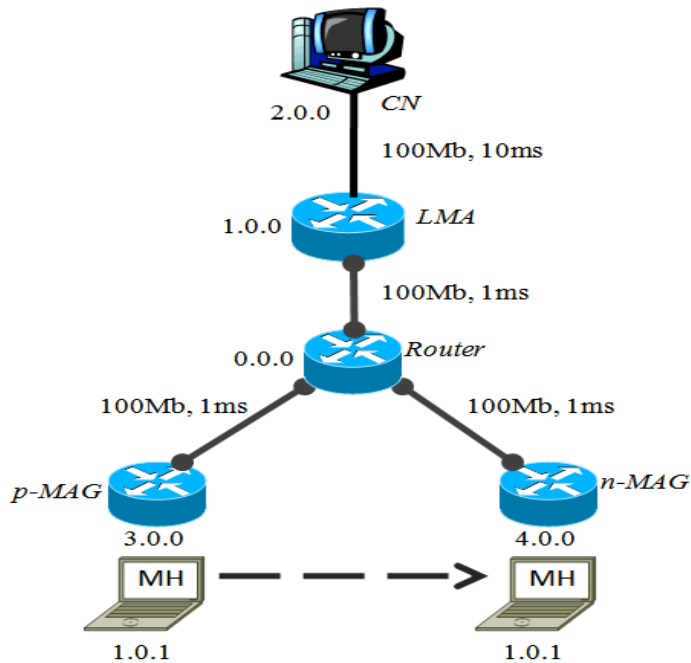

Figure 8. Simulation Scenario Setup of proposed integrated solution of PMIPv6-MIH

In the above simulation scenario, the p-MAG and n-MAG are in separate subnets. The two MAGs have both $L_2$ and $L_3$ capabilities that handles handovers. The router is interconnected to the LMA by a series of agents that are organized in a hierarchical tree structure of point-to-point wired links. The router is interconnected to the LMA by a series of agents that are organized in a hierarchical tree structure of point-to-point wired links.

The packet flow of CBR and FTP are simulated and transmitted from the CN to the MH using UDP and TCP. The link delay between the CN and the LMA is set at 10ms while the bandwidth is set at 100Mb. The link delay between the LMA and the respective MAGs is set at 1ms. The CBR and FTP packet size is set at 1000 and 1040 bytes while the interval between successive packets is fixed at 0.001 seconds.

The flow of video traffic is simulated and transmitted from the CN to the MH using myUDP. The video packet size is set at 1028 bytes while the interval between successive packets is also fixed at 0.001 seconds.

## 5.2. Simulation Results 1

Simulation results for intra-domain traffics are obtained as follows:

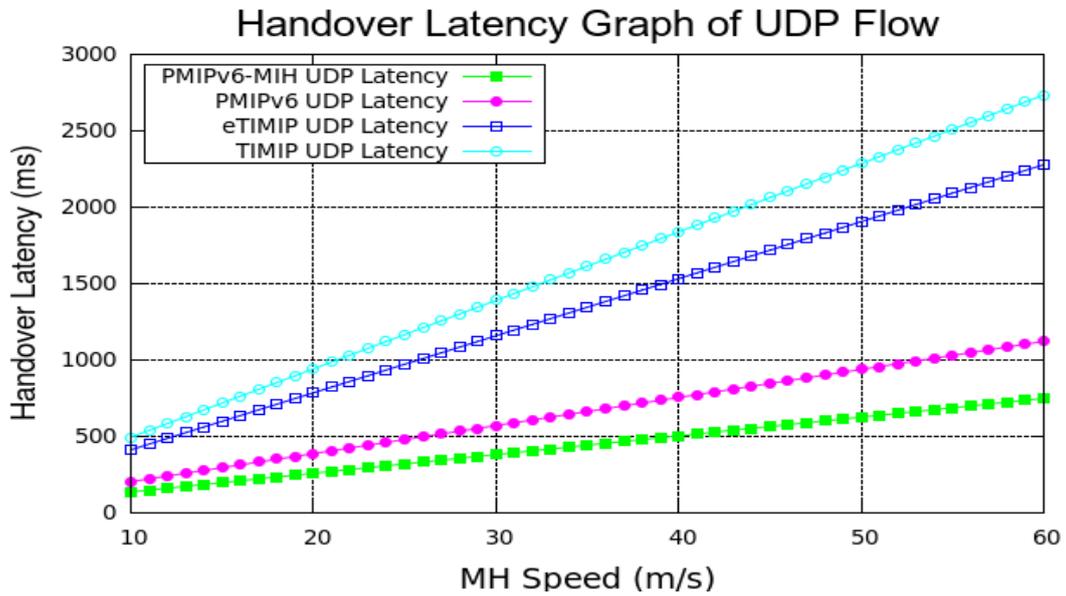

Figure 9. Handover Latency of UDP Flow in micro mobility domain

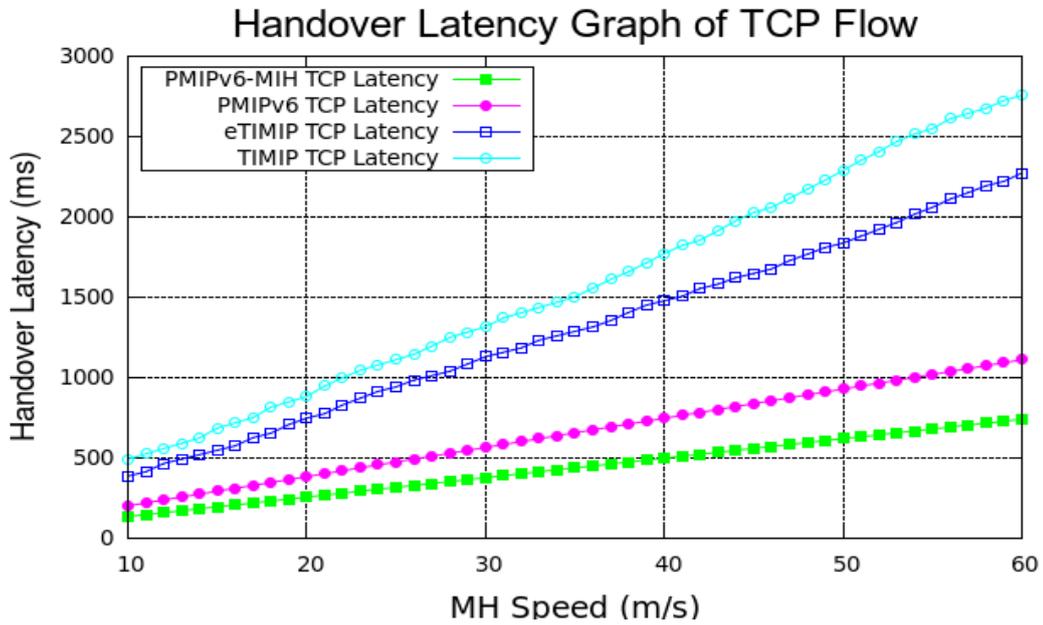

Figure 10. Handover Latency of TCP Flow in micro mobility domain

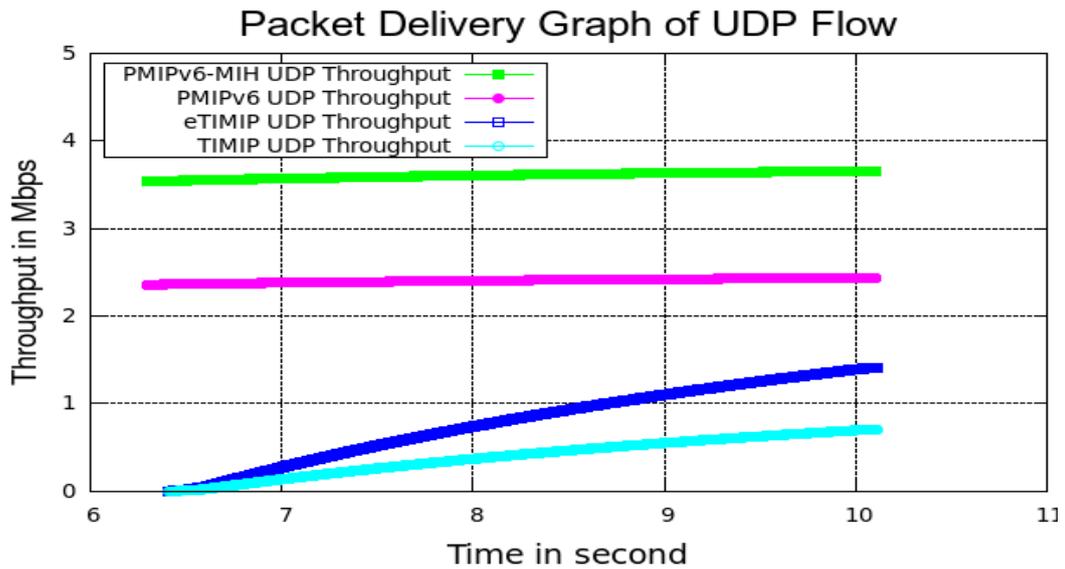

Figure 11. Throughput (Mbps) of UDP Flow in micro mobility domain

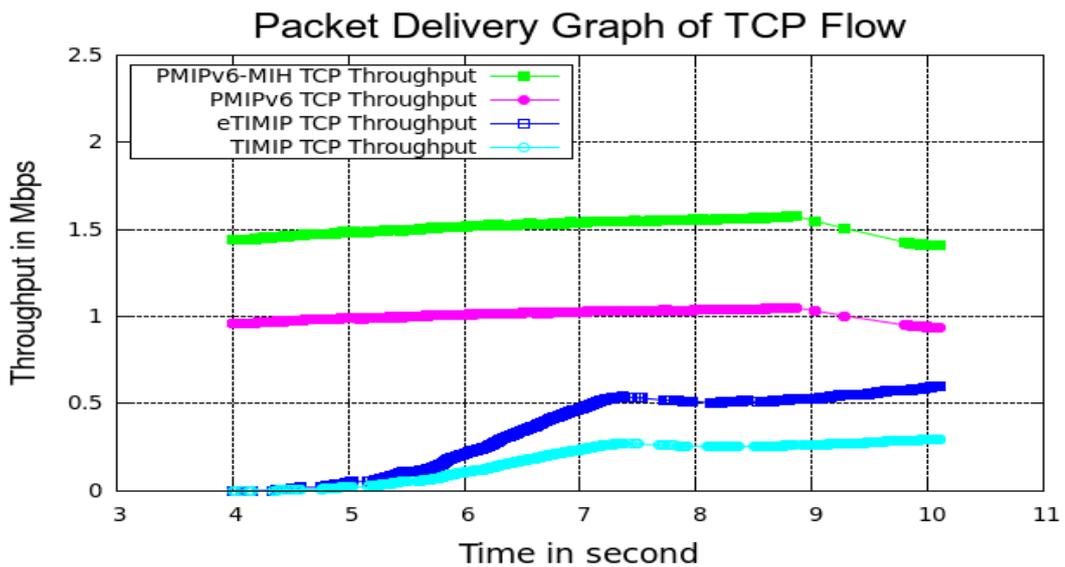

Figure 12. Throughput (Mbps) of TCP Flow in micro mobility domain

As per results shown in figures 9, 10, 11 and 12, it is observed that UDP and TCP performance of eTIMIP and TIMIP increased the handover latency during the MH moves to new network in micro mobility domain. It also noted from the simulation results that performance of throughput also shown degradation. This is due to the fact that, when MH moves away from one network to another in micro mobility domain with high speed mobility, there are lots of operations to perform between the changes of network, such as configuring new CoA, DAD operation, binding registration and MD.

In comparison to PMIPv6, it does not require CoA and DAD as MH is already roaming in the PMIPv6 domain. Once the MH has entered and is roaming inside the PMIPv6 domain, CoA is not relevant since according to the PMIPv6 specification, the MH continues to use the same address configuration. The operation of a DAD is required for a link-local address since address collision is possible between MH, MAG and all MH's attached to the same MAG. The DAD operation may significantly increase handover latency and is a very time consuming procedure. As DAD requires around one second (or even much than one sec.), PMIPv6 introduce a per-MH prefix model in which every MH is assigned a unique HNP. This approach may guarantee address uniqueness. But still PMIPv6 suffers from a lengthy handover latency and packet loss during the handover process when MH speed is high. To overcome these problems, we proposed integrated solution scheme for PMIPv6 that can send the MH-profile to the n-MAG through ND message before handover on $L_3$ and also reduce the time on $L_2$ scanning by informing the MH about the channel information of next APs using MIH services.

Based on the proposed solution scheme, the result of handover latency and throughput are better than other mobility protocols. The reason of reduce handover latency and improve throughput in micro mobility domain as below:

- ❖ The time required to obtain MH profile information can be omitted since n-MAG performs this information retrieval prior to MH's actual attachment.

- ❖ As the specification of PMIPv6, the time needed to obtain the DAD operation and configure new CoA can also be non-appreciable since n-MAG obtains MH profile and network information through the ND message and performs a pre-DAD procedure like assigning a unique HNP during available resource negotiation with p-MAG and the MH continues to use the same address configuration.

- ❖ The time required to obtain mobility-related signaling massage exchange during pre-registration may not be considered since this negotiation is established through the ND message before MH attachment. Since the MH is already pre-registered and there is no need to confirm at the n-MAG, therefore the last PBA message send from the LMA may not be considered.

### 5.3. Simulation Results 2

Simulation results of video transmission over PMIPv6 and PMIPv6-MIH in a micro mobility domain are obtained as follows:

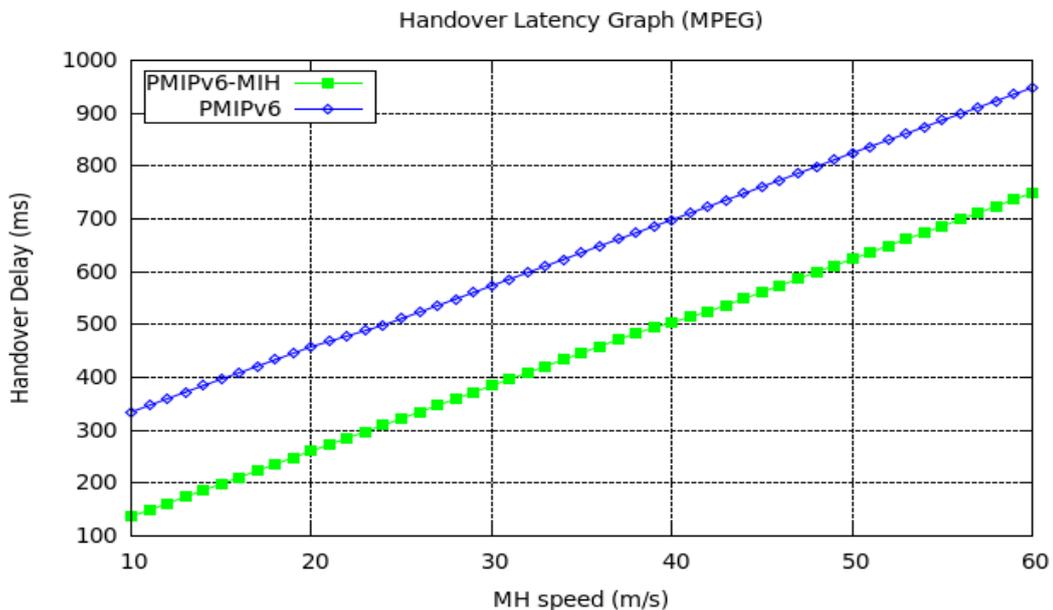

Figure 13. Latency of network-based mobility management protocols

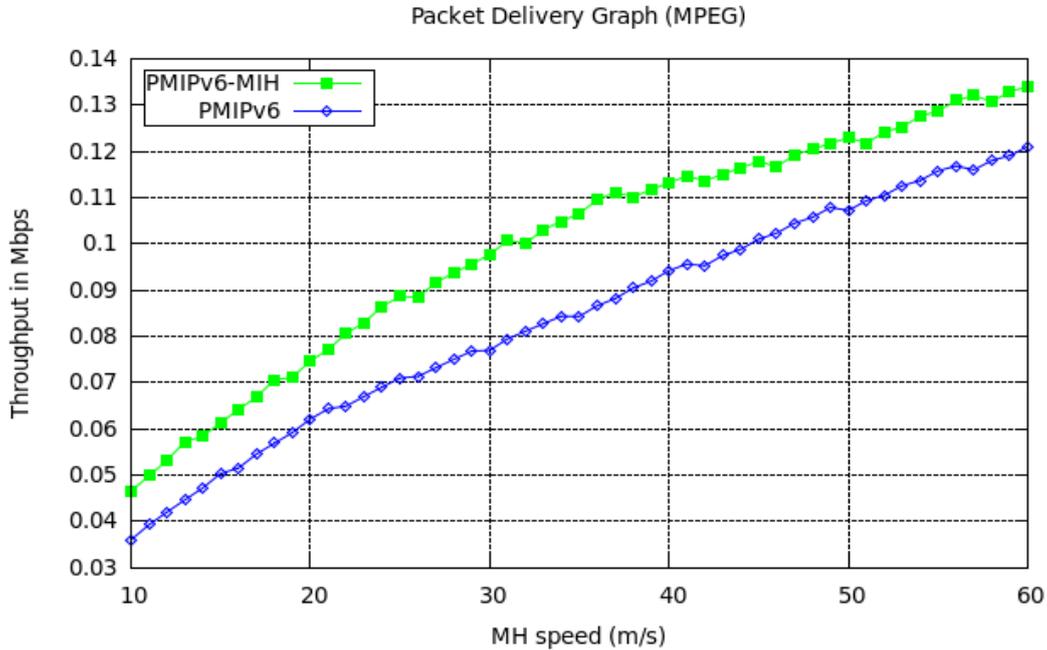

Figure 14. Throughput of network-based mobility management protocols

When the simulation is started, the video start to be transmitted from CN to MH. We can obtain how many frames and packets have been sent or lost after video has been received by MH. The results are as follows:

*PMIPv6-MIH*
```
Packet sent:p->nA:4651, p->nI:1106, p->nP:1459, p->nB:2085
Packet lost:p->lA:108, p->lI:40, p->lP:25, p->lB:43

Frame sent:f->nA:1331, f->nI:148, f->nP:296, f->nB:886
Frame lost:f->lA:38, f->lI:5, f->lP:8, f->lB:25
```

*PMIPv6*
```
Packet sent:p->nA:4651, p->nI:1106, p->nP:1459, p->nB:2085
Packet lost:p->lA:268, p->lI:88, p->lP:71, p->lB:109

Frame sent:f->nA:1331, f->nI:148, f->nP:296, f->nB:886
Frame lost:f->lA:91, f->lI:10, f->lP:20, f->lB:61
```

The results of packet and frame are as follows: A denotes number of packets or frames, I, P, B are three different types of frames of MPEG4. The cause of high packet loss, increased in handover latency and decreased throughput of PMIPv6 is due to detection of MHs' detachment and attachment events remain difficult, the need to acquire information from the policy store/AAA server, high number of packets are being transmitted to the p-MAG which cause loss many packets. In order to solve the problems of PMIPv6, the use of our proposed PMIPv6-MIH is expected to reduce lengthy handover latency, decrease the packet loss, increase the throughput and PSNR since the connection to the n-MAG is established before the MH arrived. The time required to obtain MH profile information can be omitted since p-MAG sends all information through the ND message to the n-MAG before MH's actual attachment. Eventually, RS message transmission time may not be appreciable because of the specification of PMIPv6.

The time required to obtain mobility-related signaling message exchange during pre-registration may not be considered since this negotiation is established before MH attachment. Since the MH is already pre-registered and all the necessary information already send to n-MAG through the ND message, therefore no need to confirm at the n-MAG and the last PBA message sent from the LMA may not be considered.

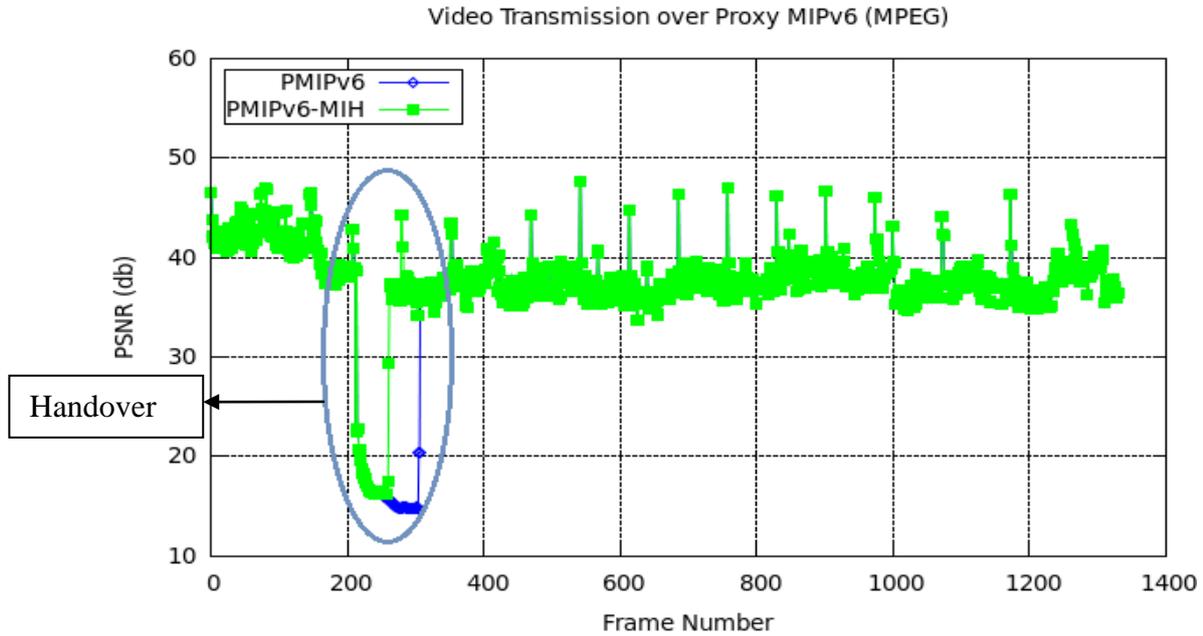

Figure 15. PSNR of network-based mobility management protocols

In figure 15, we can see at the time the PSNR falls when the 250th frame is being transmitted, which is due to the time that handover is taking place during the high speed movement of MH. In addition, the raw video and received video of network-based mobility management protocols can be differentiated with the measurement of computing PSNR [17].

$$PSNR\ (n)_{dB} = 20lg_{10}\ \frac{V_{peak}}{\sqrt{\frac{1}{N_{col}N_{row}}\sum_{i=0}^{N_{col}}\sum_{j=0}^{N_{row}}[Y_S(n,i,j)-Y_D(n,i,j)]^2}}$$

where $V_{peak} = 2^k-1$, k denotes number of bits per pixels, Y denotes component of source image S and destination image D. PSNR is widely used to compare the video quality of compressed and decompressed images.

The quality of transmitted video images is shown in table below:

*Raw Video*

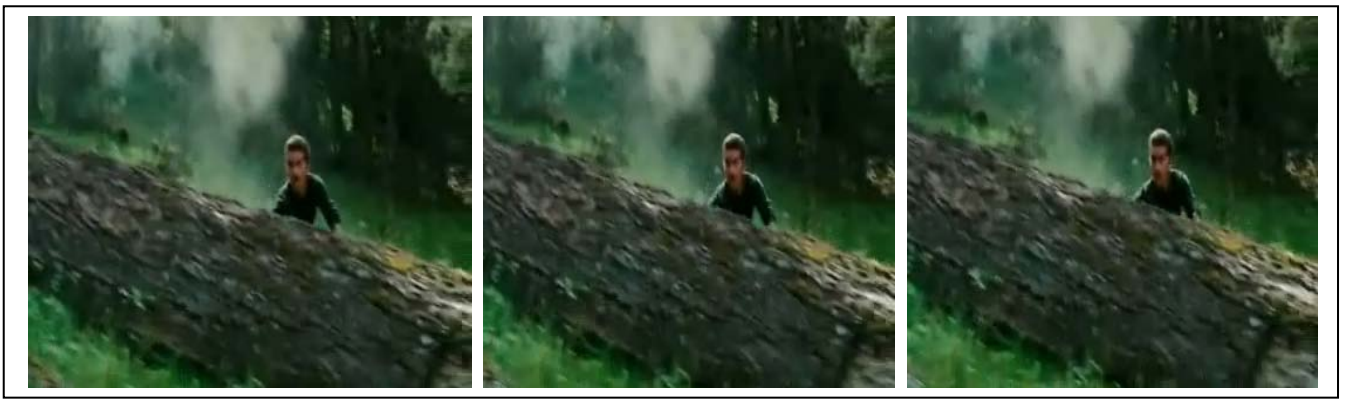

*Video received over PMIPv6*

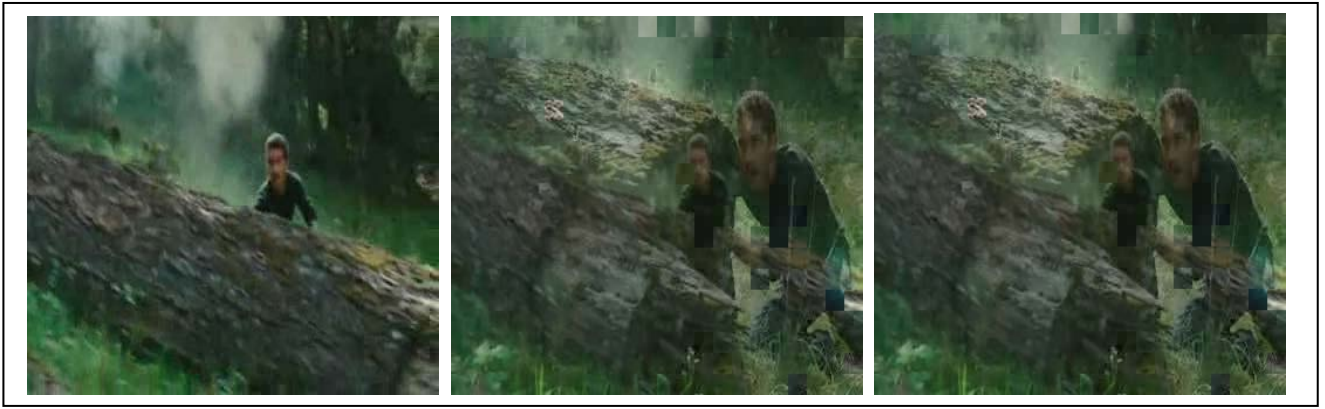

*Video received over PMIPv6-MIH*

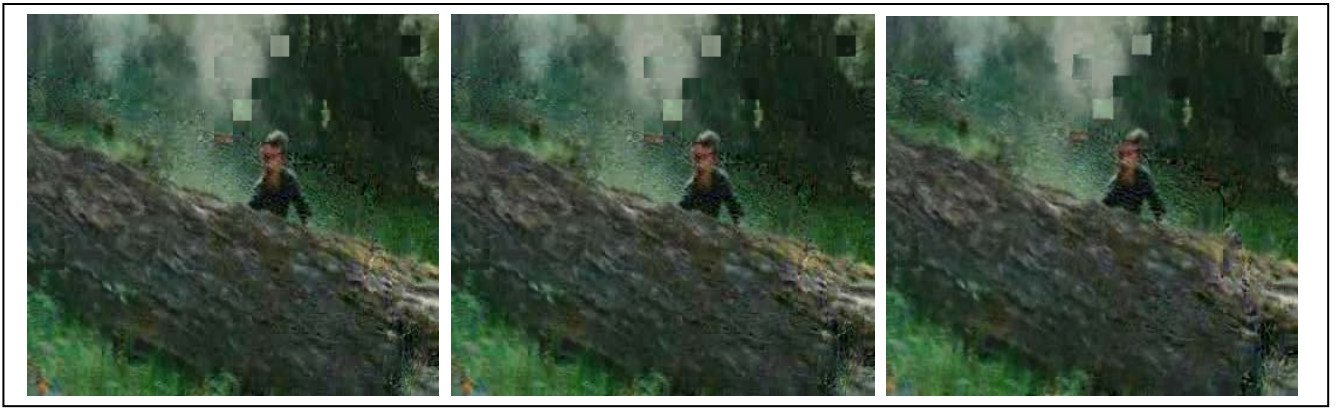

## 6. CONCLUSION

In this paper, simulations were conducted using ns-2 to evaluate, compare and examine the network-based mobility management protocols and micro mobility protocols with a high speed movement of MH for our proposed integrated solution scheme under intra-domain approaches. As for performance evaluation, we compared performance indicators such as handover latency, packet loss, throughput and video transmission quality. From our analytical analysis and simulation results, we are able to show that our proposed integrated solution of PMIPv6 (PMIPv6-MIH), as a network-based mobility management protocol, performs better than PMIPv6. For our future work, we would like to improve the packet loss, handover latency and also improve the performance of our proposed PMIPv6-MIH for real-time applications in a macro mobility domain with a high speed movement of MH.

# Biographies

**Authors**

*Md. Mahedi Hassan* received his B.S. degree in Information System Engineering from Multimedia University, Malaysia in 2008. He is a M.Sc. candidate in the Faculty of Information Technology, Multimedia University, Malaysia. His current research interests include mobile & wireless communication, high-speed network and networking.

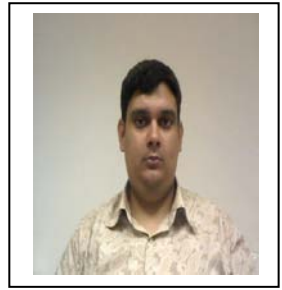

*Poo Kuan Hoong* received his B.Sc. and M.IT. degrees in 1997 and 2001 respectively from the Universiti Kebangsaan Malaysia, Malaysia, and his Ph.D. degree in 2008 from the Department of Computer Science and Engineering, Nagoya Institute of Technology, Japan. Since June 2001, he has been a lecturer in the Faculty of Information Technology at Multimedia University, Malaysia. His research is in the areas of peer-to-peer networks and distributed systems. He is a member of IEICE, IEEE and ACM.

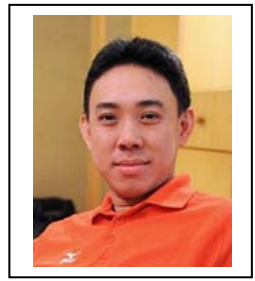